\documentclass[letter]{aa} 
\usepackage{graphicx,natbib}
\usepackage{lipsum}
\usepackage{latexsym,amsmath,tabularx,amssymb}
\usepackage{tablefootnote}
\usepackage{mathrsfs}
\bibpunct{(}{)}{;}{a}{}{,}
\usepackage[varg]{txfonts}
\include{def}

\usepackage{bm}
\usepackage{units}
\usepackage{color}
\usepackage{url}

\usepackage{ulem} 

\newcommand{\ME}{\rm{M}$_{\oplus}$} 
\newcommand{\RE}{\rm{R}_{\oplus}} 
\newcommand{\Mearth}{\text{M}_{\oplus}} 
\newcommand{\Msun}{\text{M}_{\odot} }

\newcommand{\Mcore}{M_{\text{core}}}

\newcommand{\Rp}{R_{\text{P}}}

\newcommand{\Mp}{M_{\text{P}}}
\newcommand{\Miso}{M_{\text{iso}}}

\newcommand{\fhhe}{f_{\text{HHe}} }

\newcommand{\Md}{M_{\rm d,0}}

\begin{document}

\title{ The Nature of the Radius Valley: \\Hints from Formation and Evolution Models}

 \titlerunning{The Radius Valley: Hints from Formation and Evolution}
\authorrunning{Venturini et al.}

\author{Julia Venturini \inst{1},
Octavio M. Guilera \inst{2,4,5},
Jonas Haldemann\inst{3}, 
Mar\'{\i}a P. Ronco \inst{4,5}, 
Christoph Mordasini\inst{3}}

\offprints{J. Venturini}
\institute{International Space Science Institute, Hallerstrasse 6, CH-3012 , Bern, Switzerland, 
\and
Instituto de Astrof\'{\i}sica de La Plata, CCT La Plata-CONICET-UNLP, Paseo del Bosque S/N (1900), La Plata, Argentina.
\and
{Department of Space Research \& Planetary Sciences, University of Bern, Gesellschaftsstrasse 6, CH-3012 Bern, Switzerland}
\and
Instituto de Astrof\'{\i}sica, Pontificia Universidad Cat\'olica de Chile, Santiago, Chile.
\and
Millennium Nucleus of Planetary Formation (NPF), Chile.
\\
\email{julia.venturini@issibern.ch}}

\abstract{The existence of a Radius Valley in the Kepler size distribution stands as one of the most important observational constraints to understand the origin and composition of exoplanets with radii between that of Earth and Neptune. In this work, we provide insights into the existence of the Radius Valley from, first, a pure formation point of view, and second, a combined formation-evolution model. We run global planet formation simulations including the evolution of dust by coagulation, drift and fragmentation; and the evolution of the gaseous disc by viscous accretion and photoevaporation. A planet grows from a moon-mass embryo by either silicate or icy pebble accretion, depending on its position with respect to the water ice line. We include gas accretion, type-I/II migration and photoevaporation driven mass-loss after formation. We perform an extensive parameter study evaluating a wide range in disc properties and embryo's initial location. We find that due to the change in dust properties at the water ice line, rocky cores form typically with $\sim$3 \ME \, and have a maximum mass of $\sim$5 \ME, while icy cores peak at $\sim$10 \ME, with masses lower than 5 \ME \, being scarce. When neglecting the gaseous envelope, the formed rocky and icy cores account naturally for the two peaks of the Kepler size distribution. The presence of massive envelopes yields planets more massive than $\sim$10 \ME \, with radii above 4 $\RE$. While the first peak of the Kepler size distribution is undoubtedly populated by bare rocky cores, as shown extensively in the past, the second peak can host half-rock/half-water planets with thin or non-existent H-He atmospheres, as suggested by a few previous studies. Some additional mechanism inhibiting gas accretion or promoting envelope-mass loss should operate at short orbital periods to explain the presence of $\sim$10-40 \ME \, planets falling in the second peak of the size distribution.}


\maketitle

\section{Introduction} \label{intro}

The California-Kepler Survey revealed that  
 exoplanets within a 100-day orbital period present a bimodal size distribution, with peaks at $\sim$1.3 and $\sim$2.4 $\RE$ \citep{Fulton17}. 
More recent analysis of better characterised sub-samples showed the peaks at $\sim$1.5 and $\sim$2.7 $\RE$, and the valley or gap at $\sim$1.9-2 $\RE$ \citep{VanEylen18, Martinez19, Petigura20}.   

The valley can be explained by atmospheric mass-loss mechanisms, such as photoevaporation \citep[e.g][]{Owen17, JinMord18} or core-powered mass-loss \citep[e.g][]{Ginzburg18, Gupta19}. 
Both models are able to reproduce the correct position of the valley only if the naked-cores resulting from the mass-loss are rocky in composition.  
This has led to the interpretation that most Kepler planets with radii between Earth and Neptune accreted only dry condensates and were therefore formed within the water ice line \citep{Owen17, Gupta19}.  


From a formation point of view, it is hard to envision scenarios where planets with masses below 20 \ME \, are devoid of water. 
Indeed, accretion beyond the ice line is usually prominent, and type-I migration tends to move planets in the mass range of $\sim$1-20 \ME\, inwards in a very effective way \citep[e.g.][]{Tanaka02}. 
Hence, a pure dry core composition for most short period exoplanets is not really expected from formation models \citep{Raymond18, Bitsch19a, Brugger20}.
A possible way out is to invoke migration traps due to the existence of dead zones in the disc \citep{Alessi20}. However, even if the super-Earths produced by those models are dry, they cannot account for the Kepler size bimodality.

Recent studies, based on Mass-Radius relations, suggest, on the other hand, that only the first peak of the radius distribution corresponds to rocky planets, while the second are water-rich objects \citep{Zeng19}.  
The problem with associating the second peak to water-rich planets is that it cannot explain why such planets do not fill the valley. Indeed, cores containing 50\% rock-50\% ice by mass would fall in the radius valley if they had a mass of $\sim$3 to 6 \ME \, \citep{Sotin_2007, Zeng19, haldemann_2020, Owen17, Gupta19}. 
\citet{Zeng19} showed that the Kepler size distribution can be matched if the icy planets are assumed to follow the mass distribution suggested by RV measurements, which encompasses masses in the range of $\sim$6-15 \ME, with a peak at $\sim$9 \ME. However, no explanation for the origin of such mass distribution is offered.

In an accompanying paper \citep[][hereafter Paper I]{Venturini2020a} we show that when pebble accretion is computed self-consistently from dust growth and evolution models, pure rocky planets are typically less massive than 5 \ME.
In that work, we also show that the change of dust properties at the ice line affects dramatically the growth mode of planets, which was originally proposed by \citet{Morby15} to explain the dichotomy of gaseous versus terrestrial planets in the Solar System. 
In this letter, we show that a bimodality in core mass/composition from birth naturally renders a radius valley at $\sim 1.5- 2 \, \RE$. We additionally discuss the effect of gaseous envelopes and their photoevaporation on the Kepler size bimodality.

\section{Methodology in brief}\label{sec_method}
Our physical model is the same as in Paper I, except that planets are always allowed to migrate. We recall it here briefly.
An embryo grows from lunar-mass by pebble and gas accretion, embedded in an $\alpha$-disc that undergoes X-ray photoevaporation from the central star. The adopted $\alpha$-values are $10^{-3}$ and $10^{-4}$. The pebble surface density is computed self-consistently from dust coagulation, fragmentation, drift and ice sublimation at the water ice line \citep{Birnstiel11, Drazkowska16,Guilera20}. We consider the growth of one embryo per disc, which accretes either rocky or icy pebbles, depending on its position with respect to the water ice line. The fragmentation threshold velocity of icy pebbles is taken as v$_{\rm th}=10$ m/s and v$_{\rm th}=1$ m/s for rocky ones (see Paper I and Appendix \ref{App_vth} for a discussion about this choice). `Rocky' means, in this work, Earth-like composition (i.e, 1/3 iron and 2/3 silicates by mass).

Gas accretion is computed, both in the attached and detached phases. 
To reduce computational time in the attached phase, the interior structure of the planets is calculated using the method presented in \citet{Alibert19}, which uses deep neural networks, trained on pre-computed structure models. Before the core reaches the pebble isolation mass (when $\Mcore$ = 0.9 $\Miso$), we switch to solve the internal structure equations to capture the increase of gas accretion resulting from the halt of pebble accretion (see Paper I). Type-I migration prescriptions 
account for the possibility of outwards migration due to corotation \citep{JM17} and thermal torques \citep{masset2017}. Planets switch to type-II migration once a partial gap opens in the disc \citep{Crida2006}. 
We perform in total 665 planet formation simulations, spanning a wide range in initial conditions and disc properties, as detailed in Appendix \ref{App_initialCond}.

Once the disc dissipates, the final planetary radius is computed after 5 Gyr of cooling and photoevaporation by solving the internal structure equations and checking where the semi-grey atmosphere becomes optically thick (see details in Paper I). We employ two photoevaporation models. In Model A, the water is assumed in the form of ice, mixed with the rocks in the planetary core. A H-He envelope lays on top and undergoes mass-loss. 
In Model B, the water is assumed in the form of vapour and uniformly mixed with  the H-He, conforming a H-He-H$_2$O envelope, with all its compounds affected by the mass-loss \citep[see details in Sect.2.1.2 of][]{Mordasini20}.
For the cases where we neglect the presence of the gaseous envelope, the planetary radius is computed following \citet[][see Methods]{Zeng19}, who provides a power-law mass-radius relation determined by the mass of rocks (assumed Earth-like in composition) and water. 

\section{Results}\label{sec_results}
\begin{figure}
\begin{center}
	\includegraphics[width=0.95\columnwidth]{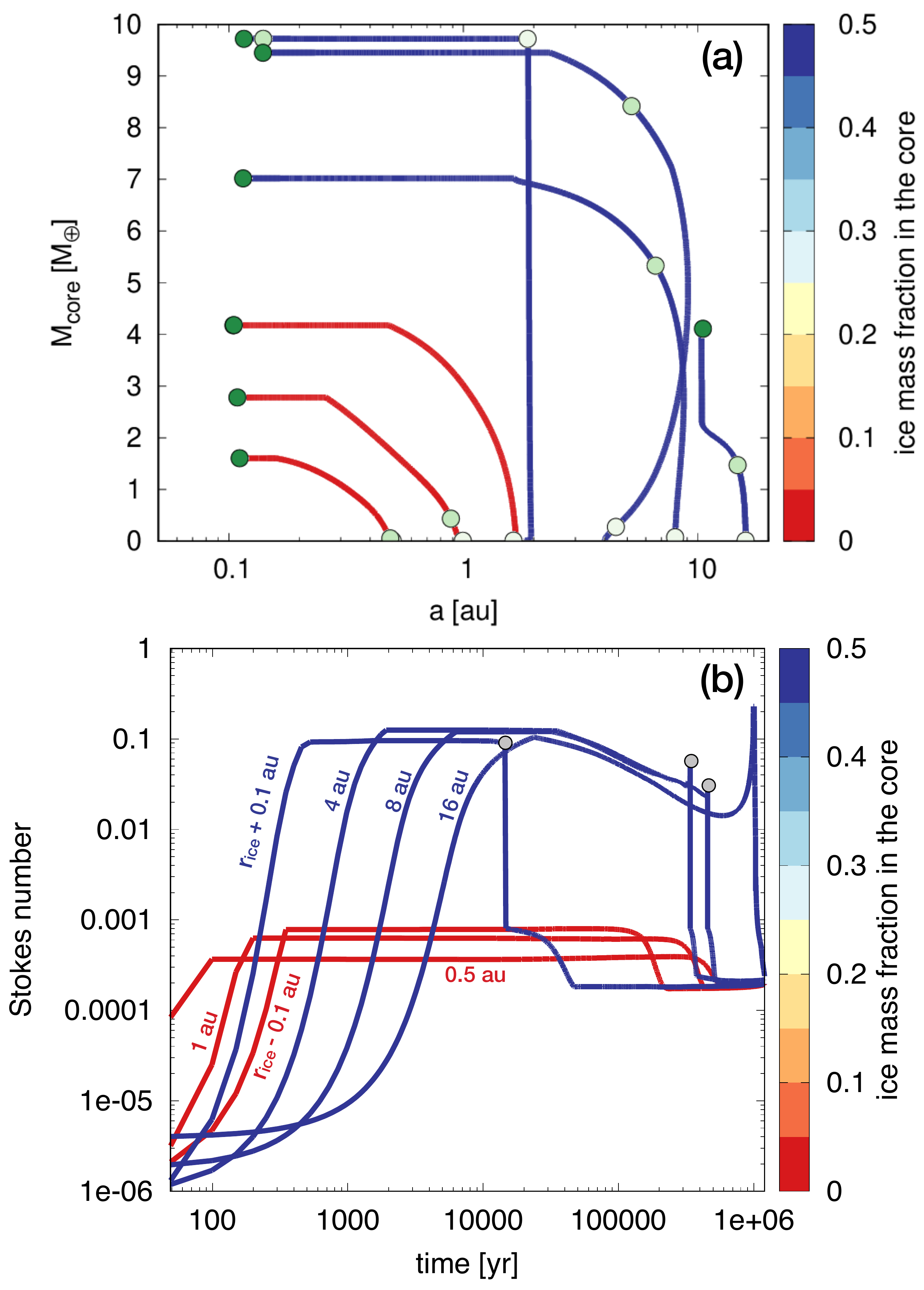}
	\caption{Top panel: formation tracks corresponding to disc 1 (see Appendix \ref{App_initialCond}), Z$_0$ = 0.0144, and $\alpha= 10^{-4}$. The white/green circles indicate the times 0.012 Myr, 0.25 Myr, and 2 Myr. (The 0.25 Myr circle of the core starting its formation just inside the ice line is below the 2 Myr circle.) $M_{\text{iso}}$ is reached in each simulation when $M_{\text{core}}$ stops growing. The core growing in a vertical line grows so fast that it practically does not migrate before reaching $\Miso$. Bottom panel: evolution of the Stokes number at the planet location for the 7 cases shown in the top panel (the labels indicate initial semi-major axis). The grey circles show the time when planets enter in the region $r<r_{\rm ice}$.} 
\label{fig_tracks}
\end{center}
\end{figure}

The water ice-line splits a protoplanetary disc into two distinct growth environments. This is because fragmentation renders silicate pebbles considerably smaller than icy ones (see Paper I),
resulting in an increase in Stokes number at the water ice-line \citep{Morby15}.
In Fig.\ref{fig_tracks}a we illustrate this effect, showing the growth tracks of seven planetary embryos that form in the same disc (one at a time). Three embryos start their growth within the ice-line and four beyond. The color-bar indicates the ice mass fraction of the core. The planets that start forming beyond the ice-line remain always water-rich ($f_{\rm ice} \approx 0.5$), because they grow fast and attain the pebble isolation mass beyond $r_{\rm ice}$ \citep[also found by][]{Brugger20, LJ14}. 
We note that all the cores that start beyond the ice-line and reach $a \lesssim 0.43$ au (or P<100 days for a Sun-mass star) are considerably more massive than the ones forming inside it. This is due to the two-order-of-magnitude jump in Stokes number (Fig.\ref{fig_tracks}b) and the fact that a large Stokes number enhances the pebble accretion rate \citep{OrmelK10, Lambrechts12}. In addition, the pebble isolation mass is larger at longer orbital periods \citep{Lambrechts14,Bitsch18, Sareh18}, which renders the icy cores effectively more massive than the rocky ones (Fig.\ref{fig_tracks}a).

\begin{figure}
\begin{center}
	\includegraphics[angle=0, width=0.95\columnwidth]{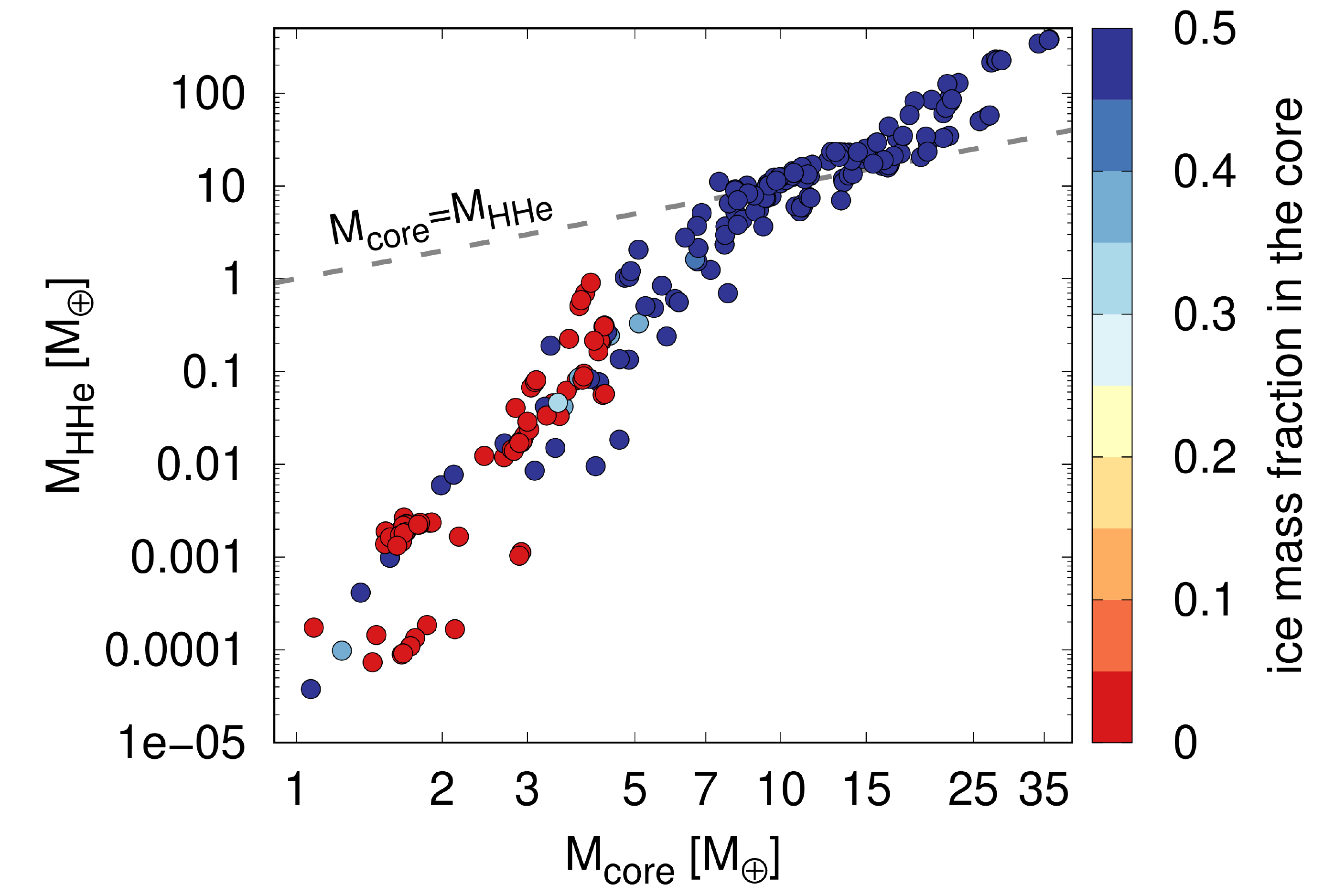}
	\caption{Core mass vs. envelope mass after formation, for all the cases with final orbital period within 100 days. }
\label{fig_McMenv}
\end{center}
\end{figure}

\begin{figure*}
\begin{center}
	\includegraphics[width=0.9\textwidth]{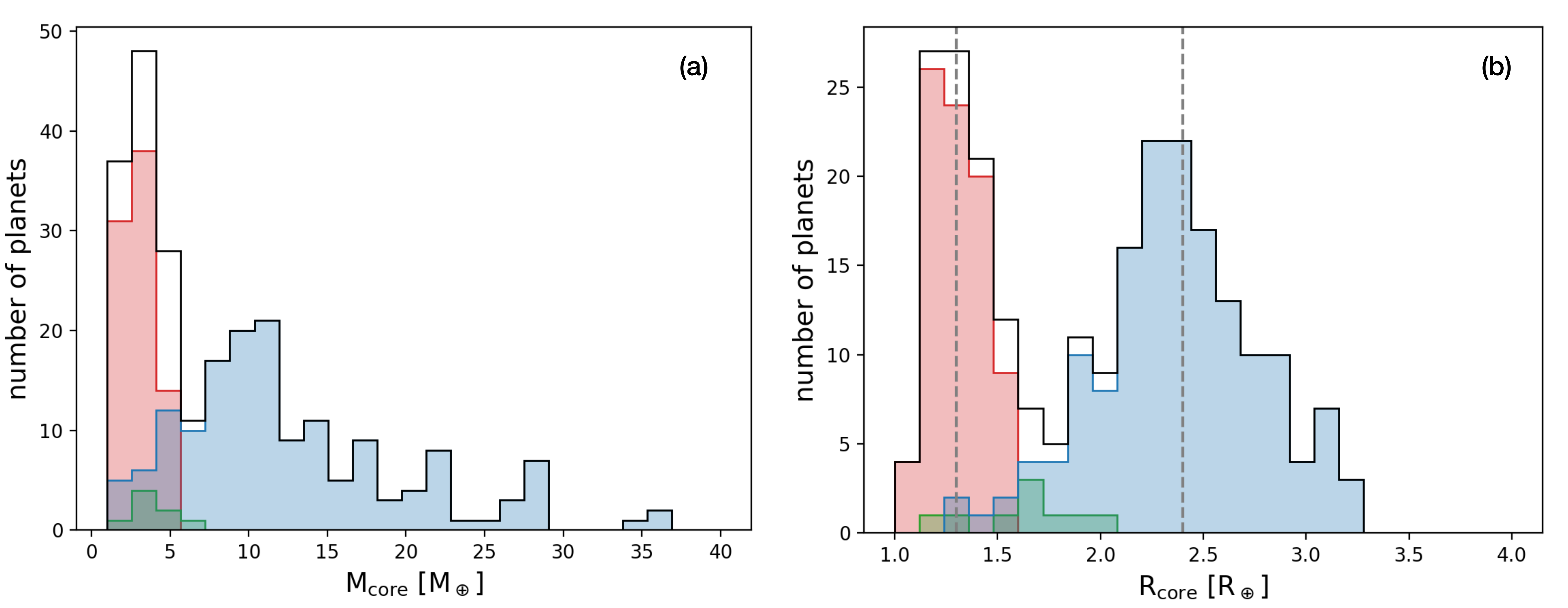}
	\caption{Histogram of core masses (left) and core radii (right) of the full population with P$\leq100$ days, just after formation. Red: $f_{\rm ice} <5\%$, green: $5\% \leq f_{\rm ice} < 45\%$, blue: $f_{\rm ice} \geq 45\%$. Black: all together.  
	The vertical lines indicate the position of the peaks as reported by \citet{Fulton17}.} 
\label{fig_hist_cores}
\end{center}
\end{figure*}

Figure \ref{fig_McMenv} shows the core and envelope mass after formation of all the simulated planets that finish with P$\leq$100 days. Again, the color-bar indicates the water mass fraction of the core. The effect of the ice-line in the core growth is noticeable: icy cores (blue) tend to be more massive than rocky ones (red). 
This is more clear when we plot a histogram of the core masses (Fig.\ref{fig_hist_cores}a). We note that the distribution of rocky core masses (f$_{\rm ice}$ = 0, red bars) is pretty narrow, with a peak at $\sim 3$ \ME \, and maximum core mass of $\sim 5$ \ME, in agreement with Paper I. On the contrary, the distribution of icy cores is more spread, with $1\lesssim\Mcore\lesssim36$ \ME. However, the peak occurs clearly for larger core masses ($\sim10 \, \Mearth$) compared to the rocky case. Indeed, the median for those planets occurs at $\Mcore$ = 10.9 \ME, and only 25\% of the icy cores have $\Mcore < 8.1 \,\Mearth$. Hence, the effect of the change of composition with the corresponding transition in the Stokes number at the water ice-line is inherited in the overall population.
Fig.\ref{fig_hist_cores}b shows the histogram of the core radii for the same cases as the left panel. Interestingly, the two peaks of the Kepler size distribution are very well reproduced, with a clear paucity of core radii at $R_{\rm valley}\approx 1.6-2 \, \RE$.

\begin{figure*}
\begin{center}
	\includegraphics[width=0.84\textwidth]{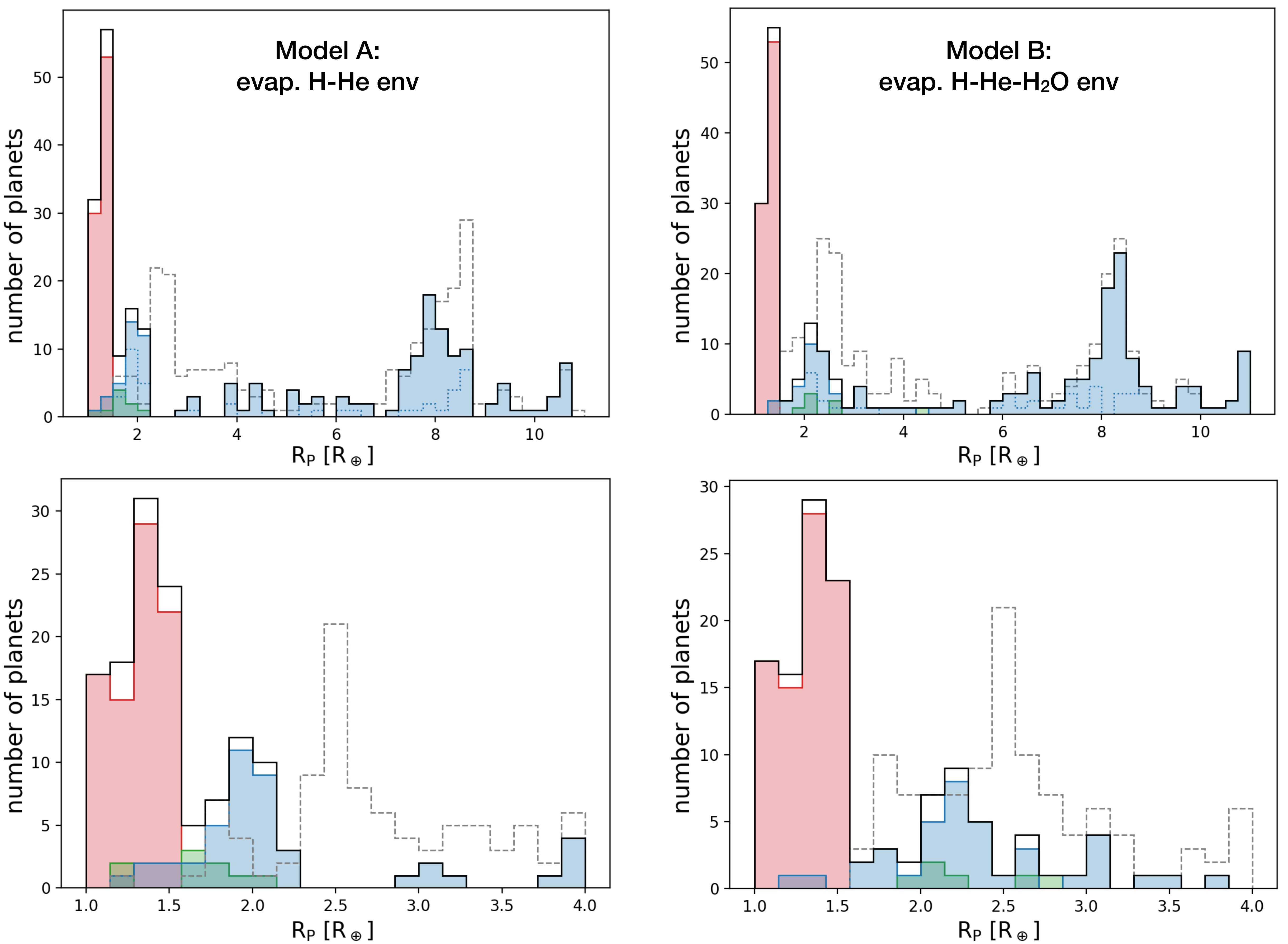}
	\caption{Radius histogram of the synthetic planets with P$\leq100$ days, after computing the cooling during 5 Gyr with mass-loss driven by evaporation (solid lines). The dashed-grey lines show the overall distribution when evaporation is neglected. Top panels: all population. Lower panels: zoom on radius between that of Earth and Neptune. Left panels: model A (evaporation of H-He envelopes). Right panels: model B (evaporation of H-He-H$_2$O envelopes). Red, blue and green indicate different initial water core fractions as in Fig.3, and black lines the overall distributions. The blue dotted-line in the upper panels shows water-rich planets born in discs of $\alpha=10^{-3}$ (the remaining cases correspond to $\alpha=10^{-4}$).}
\label{fig_histEvap}
\end{center}
\end{figure*}

However, big cores tend to accrete large amounts of gas, as Fig.\ref{fig_McMenv} shows. 
How does the size distribution look like when the gaseous envelopes are not neglected and atmospheric mass-loss is accounted for?
We show this in Fig. \ref{fig_histEvap}. Solid lines indicate size distributions accounting for photoevaporation, while dashed-grey lines show how the distributions would look like in the absence of it, to asses the precise effect of this mass-loss mechanism. We note that the appearance of the first peak at $\Rp\approx1.3\,\RE$ is a consequence of photoevaporation.
The left panels correspond to Evaporation Model A, where only the loss of H-He is considered. 
The right panels correspond to Evaporation model B, where the water is assumed to be homogeneously mixed with the primordial H-He envelope and can also be removed.
We note in this figure that the second peak (of originally icy cores) gets considerably wiped out compared to Fig.\ref{fig_hist_cores}b. Indeed, most cores of 10 \ME \, have envelopes of equal mass just after formation (Fig.\ref{fig_McMenv}), and evaporation cannot remove much gas for such massive cores. 
Then, part of the second peak moves to $\Rp \approx 8\, \RE$. Planets concentrated at this radius correspond to discs of low viscosity ($\alpha=10^{-4}$). This can be noted by comparing the solid-black and blue-dotted lines in the upper histograms of Fig.\ref{fig_histEvap}. Such low viscosity is necessary to form rocky planets (see Paper I), but creates an over-density of icy/gas-rich planets at $\Rp \approx 8\, \RE$. This could suggest a viscosity transition at the water ice-line, although $\alpha$ is expected to decrease with radial distance \citep{kretke07}. Alternatively, an efficient envelope-loss or gas-accretion-inhibitor process might operate, which renders the planets as nude cores, as Fig.\ref{fig_hist_cores}b suggests.
Despite the reduction of the amount of planets at the second peak in Fig.\ref{fig_histEvap} compared to Fig.\ref{fig_hist_cores}, it is interesting to note that
i) the paucity of planets at $\Rp \sim 1.6-2 \, \RE$, compared to $\Rp \sim 1-1.6 \, \RE$ remains for both evaporation models.
ii) for model B, a valley and small second peak appear at the position reported by \citet{Fulton17} (Fig. \ref{fig_histEvap}, right lower panel).

\begin{figure*}
\begin{center}
    \sidecaption
	\includegraphics[width=0.62\textwidth]{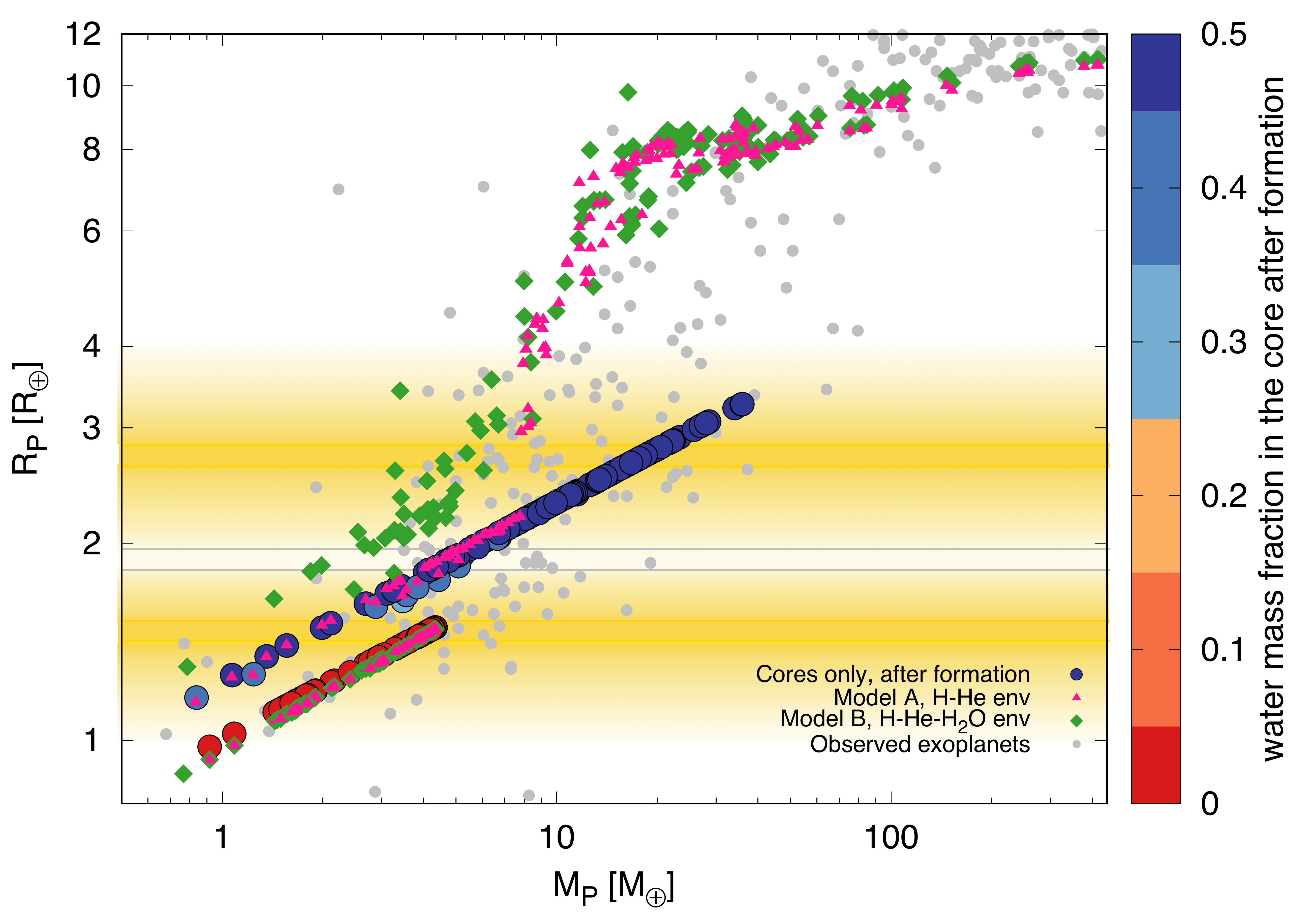}
	\caption{Mass-radius of all the planets with final orbital period within 100 days. Filled circles with color (indicating the water mass fraction of the core after formation) correspond to the mass-radius of the cores of the planets (i,e, the envelope is neglected). The radius is calculated following \citet{Zeng19} for this case. Magenta triangles show the results of evaporation of H-He after formation. Green diamonds show the same but assuming mass-loss of H, He and H$_2$O. Grey small circles are true exoplanets with orbital periods within 100 days, planet radius below 12 $\RE$, error in radius of less than 20\% and error in mass of less than 75\% (taken from the NASA Exoplanet Archive, July, 14th, 2020). Yellow shaded areas highlight the two-modes of the Kepler size distribution, with darker tones towards the peaks. The gap is delimited by the grey-horizontal lines for $1.82 \leq \Rp \leq 1.96$, following \citet{Martinez19}.}
\label{fig_MR}
\end{center}
\end{figure*}

Next, we analyse the resulting planet mass. We plot the three cases described above (bare cores after formation, and evaporation models A and B) in a Mass-Radius diagram in Fig.~\ref{fig_MR}. The bare cores are shown color-coded with the core water mass fraction. Planets run under model A are depicted as magenta triangles and as green diamonds under model B. The grey dots represent real exoplanets from the NASA Exoplanet Archive\footnote{The data was downloaded the 14th of July 2020.}. 
Yellow shaded areas highlight the two-modes of the Kepler size distribution, with darker tones towards the peaks. The gap is marked with grey lines for $1.82 \leq \Rp \leq 1.96$ following \citet{Martinez19}.

It is interesting to note that the three models overlap with existing exoplanets, and actually bracket the observed population fairly well.
Regarding evaporation model A, we note that it can strip out H-He envelopes completely for $\Mcore \lesssim 8 \, \Mearth$. Larger cores retain sufficient H-He to be kicked out of the second peak. 
Evaporation model B retains more planets in the second peak, but leaves all planets having $\Rp<4 \,\RE$ with $\Mp<6$ \ME. We discuss the implications of this in Sect.\ref{sec_discussion}.

\section{Composition of super-Earths/sub-Neptunes}
While the composition of first-peak exoplanets is undoubtedly rocky \citep[][and this work]{Owen17, JinMord18, Gupta19}, planets with radius in the second peak have an intrinsic degenerate composition, with rocky planets with thin H-He atmospheres yielding the same radius as icy-dominated objects \citep[e.g.][]{Dorn17a,Zeng19}.
Atmospheric mass-loss models tend to suggest that second-peak planets correspond to the first type. What do our combined formation and evolution models show?
We do not form rocky planets with masses above $\sim$ 5 \ME, and Model A strips the envelopes of those completely for cases with orbital periods concentrated at 10 days.\footnote{Due to our choice of disc inner edge, most of the short period planets that we form finish with $a = 0.1$ au (or P $\approx$ 10 days). We discuss this choice in Appendix \ref{AppRin}.} (At larger orbital periods some H-He can survive, see Appendix \ref{AppRin} and Paper I).
Since water is not removed in Model A, the few planets falling in the valley/second peak of that case are bare ice-rich cores (Fig.\ref{fig_MR}).

\begin{figure*}
\begin{center}
    \sidecaption
	\includegraphics[width=0.95\textwidth]{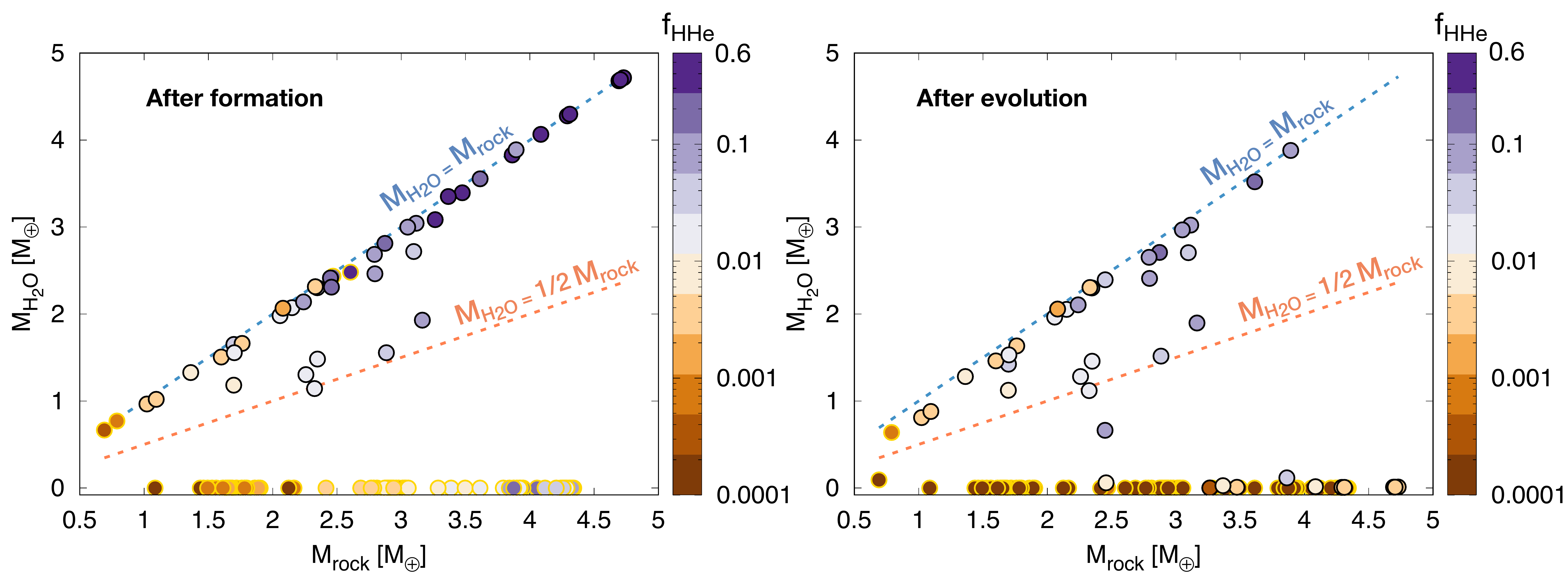}
	\caption{Bulk water versus rock content after formation (left) and after evolution (right) for Model B. The color-bar indicates the planet H-He mass fraction at each corresponding epoch. Yellow-line circles represent cases that end up in the first peak ($1< \Rp \leq 1.7$ $\RE$) and black-line circles cases that finish in the second peak (1.7<$\Rp$<4 $\RE$).}
\label{fig_comp_evapB}
\end{center}
\end{figure*}

To understand the composition of second-peak planets coming from Model B, we plot in Fig.\ref{fig_comp_evapB}, the bulk content of water and rocks, and the planet's H-He mass fraction ($\fhhe$) just after formation (left panel) and after atmospheric mass-loss by evaporation (right panel). The only quantity that remains invariable between the two panels is the mass of rocks. 
The color of the circles' border distinguishes between cases that end up in the first (yellow) or second peak (black).
Let us analyse first the case after evolution. 
First-peak planets are basically devoid of water and H-He. Regarding the second peak, most planets have water in similar amounts than rocks. These planets are not completely depleted of H-He, and have $\fhhe$ spanning 0.2\% and 10\%.\footnote{This also explains why the second small peak shown in the bottom panels of Fig.\ref{fig_histEvap} occurs at a bit larger radius for model B compared to A.} Nevertheless, a few second-peak objects are basically dry and have also a H-He mass fraction below 10\%, as found by pure evaporation models.

It is also interesting to know if first/second-peak planets accreted from inside or outside the ice-line. The left panel of Fig.\ref{fig_comp_evapB} shows the same quantities as the right one, but just after formation, before mass-loss takes place. The circles' borders still indicate the posterior belonging to the first/second peak. We note that in this case where the semi-major axis is typically $a\approx 0.1$ au (see App.\ref{AppRin}), all second-peak objects were born water-rich (also clear from the lower-right panel of Fig.\ref{fig_histEvap}), that is, they migrated from beyond the ice-line. Interestingly, even though most first-peak planets were born dry (i.e, within the ice-line), a few also started with water that was then lost.   
This means that bare rocky cores could also originate beyond the ice-line and lose all their volatile content (H, He and water) due to the stellar irradiation. The amount of first-peak objects with this origin should decrease with increasing orbital period. 

\section{Discussion}\label{sec_discussion}
We found that for the Radius Valley to exist, it is not mandatory that all planets are dry, as pure evolution models suggest. From a planet formation perspective, many of the existing super-Earths/sub-Neptunes are expected to form beyond the water ice-line, as shown in this work and many others \citep[e.g][]{Alibert2013,Bitsch19a,Schlecker20}. 
Our results indicate that second-peak planets can be often half-water/half-rock with $\sim$0.01-10\% H-He by mass (Fig. \ref{fig_comp_evapB}).
Indeed, some second-peak exoplanets present water signatures in their spectra \citep{Kreidberg20, Benneke19}. 
In addition, planets in the first peak could actually have lost all their H-He and water, and remain as bare rocky cores. Thus, planets starting their formation beyond the ice-line can end up as purely rocky as well. Our study suggest that interpreting the origin of super-Earths/sub-Neptunes can be more cumbersome than previously thought.

When analysing the final mass-radius in Fig.\ref{fig_MR}, we note that the results of our models encompass the short-period exoplanet population. When combining formation and evaporation models, it seems difficult to obtain planets with mass of $\sim$10-40 \ME \, and radius below Neptune. Nevertheless, such objects could be bare cores of half-water/half-rock if some missing mechanism could inhibit gas accretion or remove the gas after the formation. A process proposed to hinder the entire build-up of the envelope at short orbital periods is the `atmospheric recycling' \citep{Ormel15}, although more recent works adopting non-isothermal discs report that the process only abates gas accretion \citep{Kurokawa18, LambrechtsLega17, Cimerman2017}. More work on the topic is need to elucidate the importance of this mechanism. 
Another possibility is the accretion of planetesimals in addition to pebbles \citep[][]{A18, Venturini20}. In such `hybrid scenario' the heat released by planetesimals delays the accretion of gas once pebble accretion stops at isolation mass \citep{Guilera20}. 
Finally, we have neglected the effect of collisions, which can also remove gas, especially once the disc dissipates. We estimate the magnitude of collisions on the envelope-loss in Appendix \ref{AppC}. 
When one giant impact (per planet) takes place after disc dispersal, we find that: i) the Mass-Radius of the observed exoplanets is much better reproduced (Fig.\ref{fig1_AppD}) ii) the Kepler size bimodality is fairly well recovered. (Fig.\ref{fig2_AppD}a). 
Too many collisions would promote compositional mixing \citep{Raymond18}, smearing out the Radius Valley \citep{Schlecker20, VanEylen18}.

\section{Conclusions}
By studying pebble-based planet formation we found that the change of dust properties at the water ice-line combined with the increase of the pebble isolation mass with orbital distance renders two distinct populations of planetary cores, one rocky peaked at $\sim$3 \ME\, with all masses below $\sim$5 \ME, and another icy, more spread and peaked at $\sim$10 \ME. Remarkably, when neglecting the presence of the gaseous envelopes, such mass-bimodality accounts naturally for the bimodal size distribution of the Kepler exoplanets. 

When considering the formed planets with their envelopes, by computing the photoevaporation of the accreted atmospheres, we corroborate that such process can by itself render the correct radius gap. Nevertheless, contrary to pure evaporation studies, we find that the gap separates (typically) dry from wet planets. Future atmospheric characterisation with JWST and ARIEL will be crucial to learn how water-rich/poor second-peak exoplanets are, and will provide precious constraints for planet formation and evolution models.

By considering extreme-case scenarios with and without gaseous envelope, we find that the exoplanet population is fairly well bracketed by these end-members (Figure \ref{fig_MR}). This suggest, on one hand, that reality might be in between, and on the other, that a much more effective gas-accretion-inhibitor and/or gas-loss mechanism might be at operation to explain planets with masses ranging $\sim$10-40 \ME \,and falling on the second peak. The combination of different processes such as hybrid pebble-planetesimal accretion, collisions, photoevaporation and core-powered mass-loss into a single framework might be an important venue to bridge the gap between theory and observations.

\bigskip
\small{
\textit{Acknowledgements}. We thank the anonymous referee for valuable criticism. J.V. and O.M.G. thank the ISSI Team "Ice giants: formation, evolution and link to exoplanets" for fruitful discussions.  
O.M.G. thanks ISSI Bern for their support and hospitality during a monthly stay. 
J.H.  acknowledges the Swiss  National  Science  Foundation (SNSF) for supporting research through the SNSF grant  200020\_19203.
This work has been carried out in part within the framework of the NCCR PlanetS supported by the Swiss National Science Foundation. O.M.G. is partially support by PICT 2018-0934 and PICT 2016-0053 from ANPCyT, Argentina. O.M.G. and M.P.R. acknowledge financial support from the Iniciativa Cient\'{\i}fica Milenio (ICM) via the N\'ucleo Milenio de Formaci\'on Planetaria Grant. M.P.R. acknowledges financial support provided by FONDECYT grant 3190336.
}
\bibliographystyle{aa}
\bibliography{lit_2020}


\begin{appendix}

\section{Dependence on the fragmentation velocity of grains}\label{App_vth}
The core growth by pebble accretion depends sensitively on the Stokes number \citep{Lambrechts12, LJ14}, as we mentioned in Sect.\ref{sec_results}. The Stokes number is proportional to the pebbles' mean size, which is affected by the fragmentation velocity of the particles. In this work we adopted the conservative approach of considering a fragmentation threshold velocity of 1 m/s for silicate grains and 10 m/s for icy grains (Paper I). Support for these numbers stems from the experimental work of \citet{Gundlach11, Aumatell14, Gundlach15} and was adopted on the study of \citet{Drazkowska17}. 

However, recent lab experiments seem to challenge this view. \citet{Musiolik19} reported icy grains of 1.1 mm to have similar sticking properties as silicate grains for T$\lesssim$180 K and P=1.5 mbar. In the view of these new experiments, we performed a few test  cases to evaluate how our resuts depend on the adoption of more similar fragmentation threshold velocities between silicate and icy grains. 

We repeated the simulations of Fig.1 (where $v_{\rm{th}} =1$ m/s for $ r < r_{\rm ice}$ and $v_{\rm{th}} = 10$ m/s for r $\geq r_{\rm ice}$), according to the following cases:

\begin{itemize}
    \item case 1: $v_{\rm{th}} =1$ m/s along all the disc.
    \item case 2: $v_{\rm{th}} =1$ m/s for $ r < r_{\rm ice}$ and $v_{\rm{th}} = 2$ m/s for r $\geq r_{\rm ice}$.
    \item case 3: $v_{\rm{th}} =1$ m/s for $ r < r_{\rm ice}$ and $v_{\rm{th}} = 5$ m/s for r $\geq r_{\rm ice}$.
\end{itemize}

\begin{figure*}
    \centering
    \includegraphics[width=\textwidth]{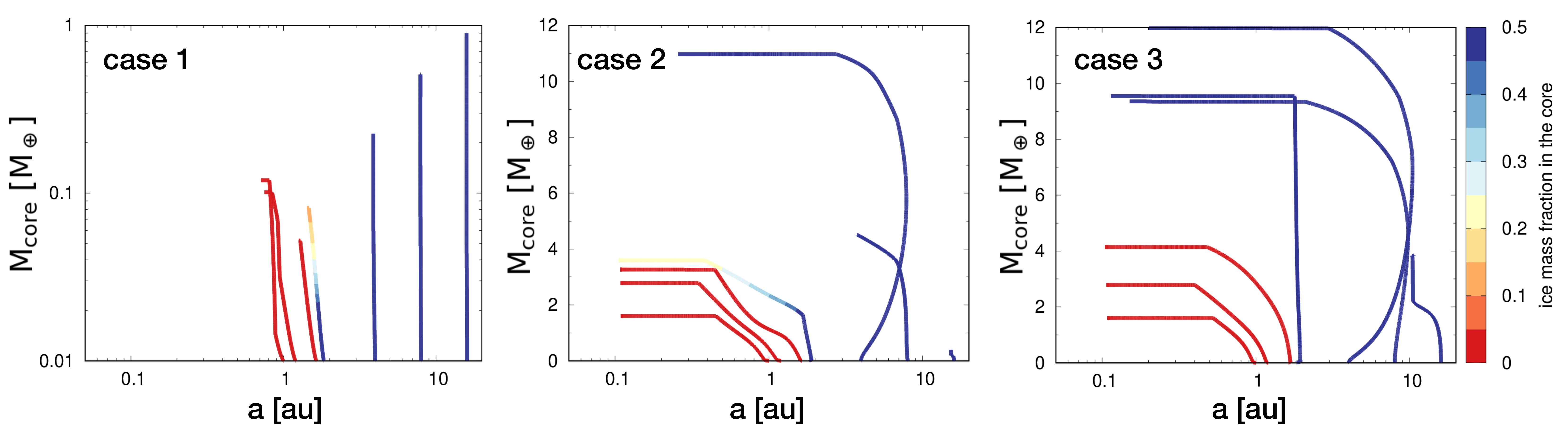}
    \caption{Repetition of Fig.1 but considering different $v_{\rm{th}}$. Case 1,2 and 3 of Appendix \ref{App_vth} are shown in the left, center and right panels; respectively. We note that, for visibility, the y-axis of case 1 has a very different scale than the other two cases. Planets of case 1 basically do not grow.}
    \label{fig_vth}
\end{figure*}
    
The results of the core growth  for these three cases are shwon in Fig.\ref{fig_vth}.
Interestingly, case 1 leads to practically no growth. The reason for this is twofold. First, the smaller particle sizes translate into a reduced drift velocity, which yields a strong reduction of the pebble flux along the entire disc. This precludes the core growth at all locations. Second, the decrease of the Stokes number resulting from the decrease of the pebble sizes, reduces the pebble accretion rate for the outer planets. This suggests that silicate and icy grains might not have exactly equivalent properties within a protoplanetary disc.
For case 2, the growth of the inner planets is very similar to Fig 1, but the growth of the outer ones is modified. There is still one icy planet that reaches $\Mcore =10 \, \Mearth$ .
For case 3, the growth of the planets is very similar to the nominal case shown in Fig.1. This suggest that a reduction of the fragmentation velocity of icy grains by a factor of two would not affect our conclusions.

Regarding the water mass fraction of the resulting cores, we note that the decrease in the pebble accretion rate promoted by a lower $v_{\rm{th}}$, makes the embryo starting its growth just outside the ice line to finish with a more mixed composition in cases 1 and 2. For case 3 the dichotomy in the resulting core ice fraction shown in Fig.\ref{fig_tracks} is recovered. 
Some previous works \citep{Bitsch19a, Schoonenberg19} find that planets starting their formation beyond the ice line continue to accrete dry pebbles within it. For this to occur, the timescale of core growth has to be longer than the timescale of migration, making it possible for the protoplanet to migrate substantially (crossing the ice line) before reaching the pebble isolation mass. This is the case for the embryo starting the formation just outside the ice line in our case 2. We remark that even reducing $v_{\rm{th}}$ of icy pebbles by a factor of 5 (case 2), the dichotomy between pure rocky and half-rock/half-ice cores is maintained unless the embryo starts to form extremely close to the ice line.

A curious aspect about the formation tracks of Fig.\ref{fig_vth} and Fig.\ref{fig_tracks} is that for case 3 and also for the nominal case (Fig.\ref{fig_tracks}), the planet starting its formation just outside the ice line grows so fast that migration does not have time to modify the trajectory, leading to an in situ formation until the attainment of $\Miso$. The planets starting their formation farther out ($a_{\rm ini} \geq 4$ au) grow slower and the torques have time to act, moving the planets typically farther away at the beginning of the growth due to the  thermal torque. The reason for the case with $a_{\rm ini} = r_{\rm ice,0} + 0.1$ au to experience this extremely fast growth (see green-white dots of Fig.\ref{fig_tracks}a) is the following. The change in $v_{\rm{th}}$ at the ice line leads to a change in pebble size and thus to a change in the drift velocities. This provokes a traffic jam in the vicinity of the ice line (at approximately $r_{\rm ice} \pm 0.5$ au), which, at early times, increases the surface density of pebbles at that location (see bottom-right panel of Fig.~1, Paper I), leading to a rise in the pebble accretion rate.

To close this section, it is important to mention that the results of \citet{Musiolik19} have been regarded as controversial by some authors. \citet{GarciaGon2020} point out that the results of \citet{Musiolik19} disagree with tensil strenght computed numerically by \citet{Tatsuuma19}. \citet{Okuzumi19} mention that Musiolik's recent experiments are inconsistent with earlier ones performed by \citet{Gundlach15} which showed efficient sticking of H$_2$O grains for temperatures down to 100 K. In addition, missing key aspects such as porosity \citep{GarciaGon2020, Krijt16} and the lack of experiments involving mixtures of silicates and ices \citep{Choukroun20} might influence the fragmentation velocities. Indeed, the very recent work of \citet{Haack20}, finds tensile strengths of mixtures of silicate and ice at 150 K being lower than previously reported. 
More experimental work is needed to pin down realistic values of dust properties under the environmental conditions of a protoplanetary disc.

\section{Disc parameters and initial conditions}\label{App_initialCond}
\begin{table*}
\caption{Observed discs from \citet{Andrews10} with their parameters and corresponding lifetimes and initial ice-line positions.}
\begin{center}
\begin{tabular}{|c|c|c|c|c|c|c|c|}
\cline{1-8}
Disk & $\gamma$ & $\Md$ [M$_{\odot}$] & r$_{\rm c}$ [au]  &\multicolumn{2}{ c| }{$\alpha=10^{-3}$} & 
\multicolumn{2}{ c| }{$\alpha=10^{-4}$} \\ \cline{5-8}
Number &  &  &  & $\tau$ [Myr] & r$_{\rm ice,0}$ [au] & $\tau$ [Myr] & r$_{\rm ice,0}$ [au] \\
\cline{1-8}
1 & 0.9 & 0.029 & 46.0 & 1.73 & 2.47 & 3.54 & 1.74    \\
2 & 0.9 & 0.117 & 127.0 & 7.24 & 2.79 & 8.84 & 1.87 \\   
3 & 0.7 & 0.143 & 198.0 & 9.08 & 1.89 & 11.07 & 1.53 \\
4 & 0.4 & 0.028 & 126.0 & 2.03 & 1.38 & 3.16 & 1.37 \\
5 & 0.9 & 0.136 & 80.0 & 7.62 & 3.79 & 9.63 & 2.30 \\
6 & 1.0 & 0.077 & 153.0 & 4.93 & 2.47 & 6.47 & 1.75 \\
7 & 0.8 & 0.029 & 33.0 & 1.61 & 2.68 & 3.65 & 1.81 \\
8 & 0.8 & 0.004 & 20.0 & 0.39 & 1.66 & 1.75 & 1.47 \\
9 & 1.0 & 0.012 & 26.0 & 0.80 & 2.30 & 3.25 & 1.69 \\
10 & 1.1 & 0.007 & 26.0 & 0.59 & 2.01 & 3.00 & 1.59 \\
11 & 1.1 & 0.007 & 38.0 & 0.56 & 1.84 & 2.87 & 1.53 \\
12 & 0.8 & 0.011 & 14.0 & 0.78 & 2.65 & 3.51 & 1.81 \\
\cline{1-8}
\end{tabular}
\end{center}
\label{TabAndrews}
\end{table*}

\begin{table*}
\caption{Adopted initial dust-to-gas ratio or disc metallicity ($Z_0$) and the corresponding [Fe/H].
[Fe/H] = log$_{10}(Z_0 /Z_{\odot})$, where $Z_{\odot}=0.0153$ is taken from \citet{Lodders09}.}
    \centering
  
    \begin{tabular}{|c | c| }
        \hline
        $Z_0$ & [Fe/H]  \\
        \hline 
        0.0068 & -0.350 \\
        0.0099 & -0.185 \\
        0.0144 & -0.025 \\
        0.0210 & 0.138 \\
        0.0305 & 0.300 \\
        \hline
        \end{tabular}
    
    \label{tabZ0}
\end{table*}

We adopt the initial gas surface density profile inferred from the observations of \citet{Andrews10}:
\begin{eqnarray}
  \Sigma_{\text{g}} &=& \Sigma_{\text{g}}^0 \left( \frac{r}{r_\text{c}} \right)^{-\gamma} e^{-(r/r_\text{c})^{2-\gamma}}, \label{eq1-sec2-0}
\end{eqnarray}
where $\Sigma_{\text{g}}^0$ is a normalisation parameter determined by the disc initial mass ($\Md$), $\gamma$ is the exponent
that represents the surface density gradient and $r_\text{c}$ is the characteristic radius of the disc. All the disc parameters are taken from \citet{Andrews10} and are shown in Table \ref{TabAndrews}, with their corresponding lifetime ($\tau$) and initial ice-line position ($r_{\rm ice,0}$). For the viscosity we consider $\alpha = 10^{-3}$ and $\alpha = 10^{-4}$. Only the low-alpha case produces pure rocky planets, as found in Paper I.

We run simulations for all the discs with lifetimes between 1 and 12 Myr (19 discs), for which we consider the initial dust-to-gas ratios (Z$_0$) shown in Tab.\ref{tabZ0}. Such wide range in dust-to-gas ratios or metallicities spans the metallicities of planet-host-stars \citep{Petigura18}. We launch 7 embryos per disc (one embryo at a time), with initial semi-major axes of $a_{\rm ini}$ = 0.5, 1, $r_{\rm ice,0}$-0.1, $r_{\rm ice,0}$+0.1, 4, 8 and 16 au.

All the embryos are inserted at t=0. We checked that changing this initial time to 0.1 Myr \citep[as is customary done in pebble accretion simulations, eg,][]{LJ14,Bitsch15b,Bitsch19a,Ogihara18a} barely modifies the results.

We note that the initial ice line position of all the discs with $\alpha=10^{-4}$ lies between 1.37 and 2.3 au (see table \ref{TabAndrews}). This is comparable to the locations reported by other works, such as \citet{Oka11, Bitsch15a} for high accretion rates onto the central star, of $\sim 10^{-7}-10^{-8} \Msun/{\rm yr}$. It is anyway important to note that our disc model uses the classical opacities of \citet{BL94}, suited for micrometer-size grains. Dust coagulation, especially for low disc turbulence, is expected to reduce the grain opacities \citep{Savvidou2020}, yielding to an ice line location closer to the central star. In Paper I and in this work we coupled in a self-consistent manner dust growth and evolution with pebble accretion. Future work should additionally address the difficult problem of coupling consistently grain growth with the disc's opacities.

\section{Dependence on the disc inner edge} \label{AppRin}
\begin{figure*}
    \centering
    \includegraphics[width=0.9\textwidth]{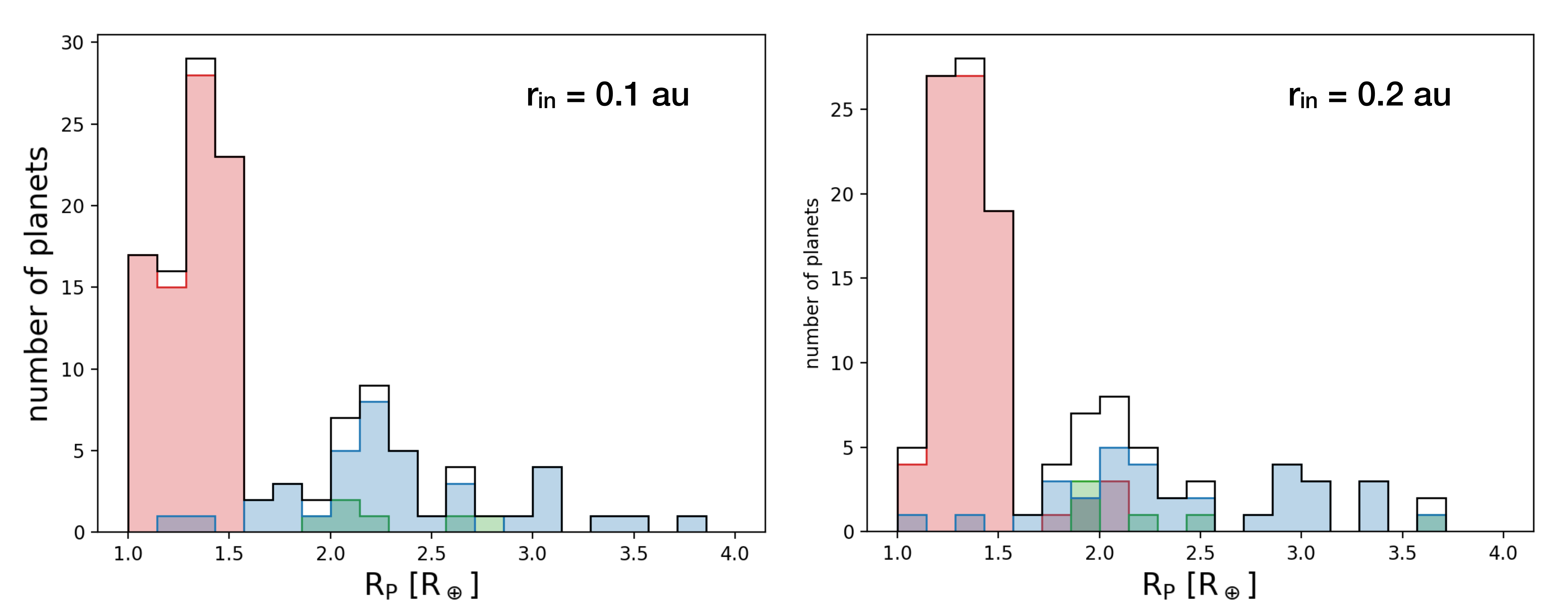}
    \caption{Same as bottom-right panel of Fig.\ref{fig_histEvap}, but comparing the nominal case  ($r_{\rm in} = 0.1$ au, left panel) with the case where $r_{\rm in} = 0.2$ au (right).}
    \label{fig_AppRin}
\end{figure*}

The inner border of the disc determines the minimum semi-major axis that planets can attain by inwards migration. When planets migrate in resonant chains, the innermost planet tends to stop its migration at or near the edge of the protoplanetary disc \citep{Cossou14}, although outwards migration can also occur due to the expansion of the inner cavity during disc dispersal \citep{Liu17}.  
Since we do not include N-body interactions nor the effect of the magnetic cavity in our calculations, most planets tend to park near the disc inner edge, assumed as $r_{\rm in} = 0.1$ au in our nominal set-up (all figures of main text). 
The final planet's position affects mainly the photoevaporation rate and hence the final mass and thickness of a planet's atmosphere. 

Our choice of nominal disc inner edge at $r_{\rm in} = 0.1$ au is based on constraints from hydrodynamical simulations \citep{Flock19}, and on the observed typical position of the the innermost exoplanet in a system \citep{Mulders18}. Still, the mean orbital period of second-peak exoplanets is $\sim$38 days \citep{Martinez19}, which corresponds to $a \approx 0.22$ au for Solar-type star. Hence, it is relevant to test how the composition of second-peak planets would change for such orbital periods.
We therefore repeat the simulations with $r_{\rm in} = 0.2$ au, together with the histograms of Fig.\ref{fig_histEvap} for Model B. Both histograms (nominal set-up and $r_{\rm in} = 0.2$ au) are shown in Fig.\ref{fig_AppRin}. 
We note that for $r_{\rm in} = 0.2$ au some planets that formed inside the water ice line (and are therefore devoid of water, the red bars) end up in the second peak, meaning that they retain some H-He atmosphere. This is a natural consequence of photoevaporation removing less gas at larger orbital distances. The longer the orbital period, the larger the amount of rocky-to-icy objects that should contribute to the second peak. Future work with population synthesis will be able to quantify this precisely and give quantitative predictions.

\section{Envelope mass-loss by giant impacts}
\label{AppC}

\begin{figure*}
\begin{center}
    \sidecaption 
	\includegraphics[width=0.65\textwidth]{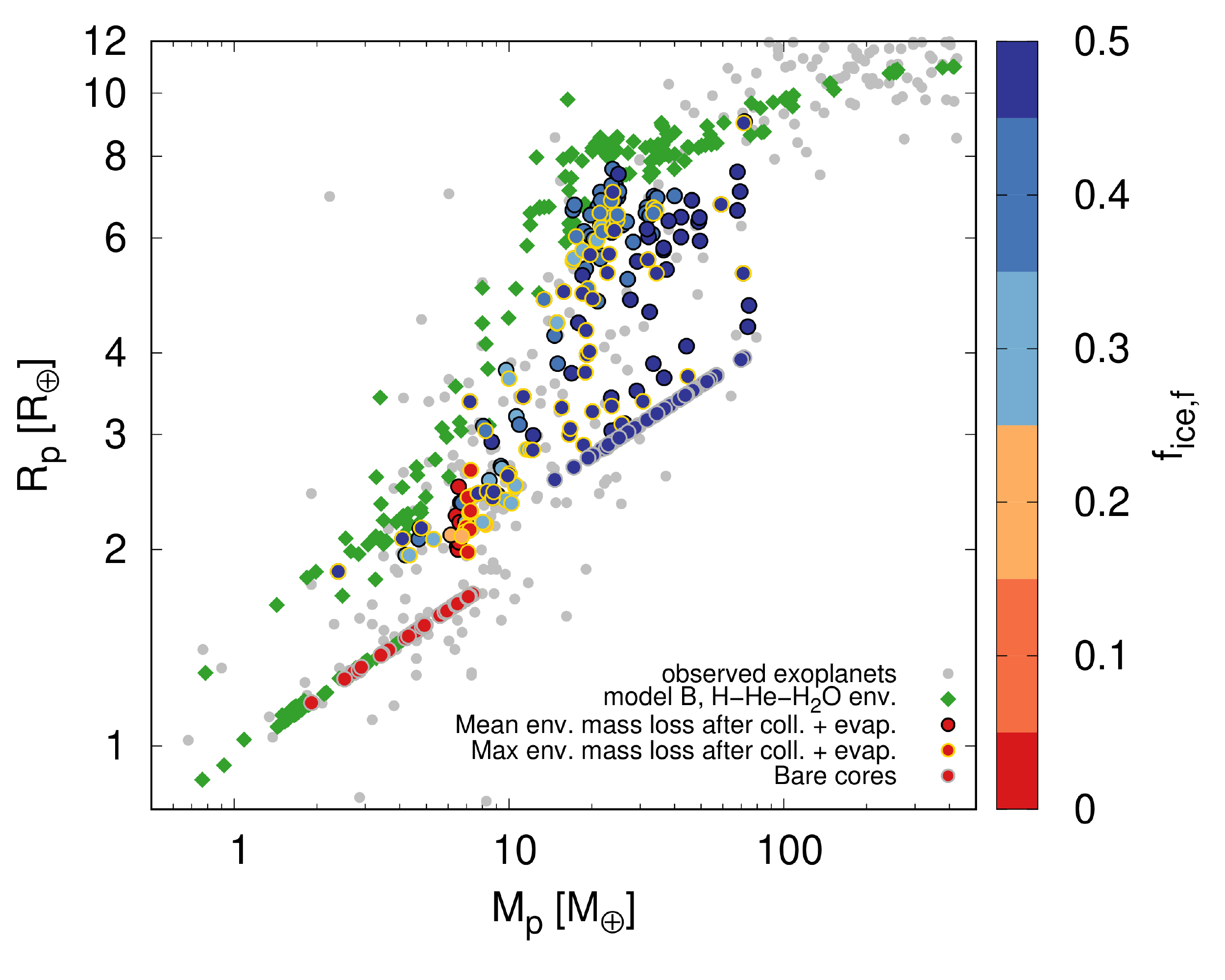}
	\caption{Same as Fig. \ref{fig_MR} but showing only model B from that figure (green diamonds), plus the results of hypothetical giant impacts followed by photoevaporation (colored circles). The color-bar represents the final water mass fraction. The grey dots represent the observed exoplanet population as in Fig.~\ref{fig_MR}.
	The yellow-border circles represent the planets that suffered the maximum envelope mass-loss due to a collision, and the black-border circles represent the mean values of envelope mass-loss for each family of collisions. 
	The grey-border circles denote the 
	bare cores that lost their envelopes completely either just after the collision or after the photoevaporation.
	}
\label{fig1_AppD}
\end{center}
\end{figure*} 

\begin{figure*}
\begin{center}
	\includegraphics[angle= 0, width=\textwidth]{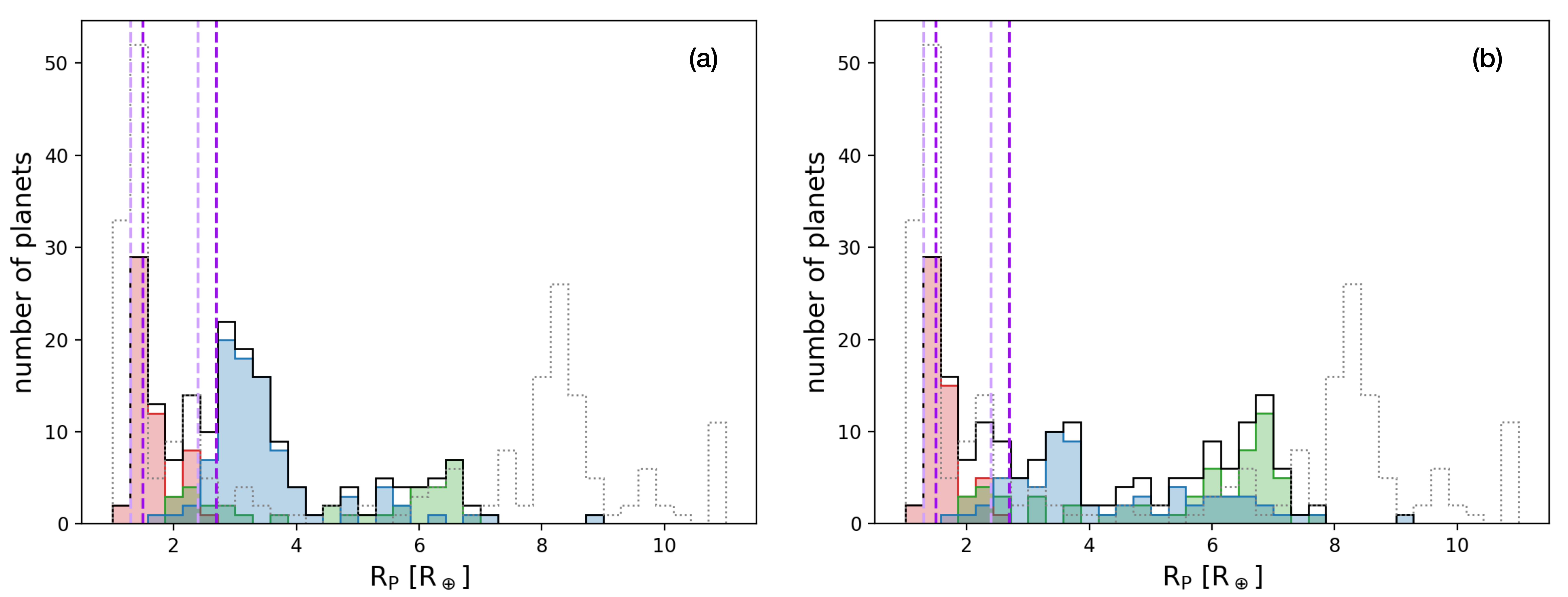}
	\caption{Size distribution of the synthetic planets after suffering one giant impact and photoevaporation. \textit{Left panel}: maximum mass removed by the collision. \textit{Right panel}: Mean mass removed by the collision. Red: $f_{\rm ice,f} <5\%$, green: $5\% \leq f_{\rm ice,f} < 45\%$, blue: $f_{\rm ice,f} \geq 45\%$, where $f_{\rm ice,f}$ is the final mass fraction of water relative to the total amount of heavy-elements. Black lines: overall size distribution. Grey-dotted lines: distribution without colissions (as in Fig.\ref{fig_histEvap}). Vertical-dashed lines: peaks of the Kepler size distribution inferred by \citet{Fulton17} (light-violet) and \citet{Martinez19} (dark-violet).}
\label{fig2_AppD}
\end{center}
\end{figure*}

Giant collisions, which may be important at the time of disc dispersal \citep{Ogihara2020b}, represent, in addition to photoevaporation, a possible mechanism that could help removing a planet's atmosphere. Although we did not consider collisions in our simulations since we formed only one planet per simulation, we can estimate in a simple way, the fraction of envelope mass-loss a planet could suffer if we allowed it to collide with another, less-massive planet, formed isolated in the same disc. 
These "hypothetical" collisions could happen within the first million years of evolution after gas dissipation and before substantial photoevaporation takes place \citep[e.g.][]{Izidoro2017}.
The goal is to compute the envelope mass-loss of a planet due to a possible collision plus the subsequent atmospheric-loss due to photoevaporation, to explore how the Mass-Radius of exoplanets (Fig. \ref{fig_MR}) and the radii distribution (Fig. \ref{fig_histEvap}) could be affected.

 Following the same procedure as in \citet{Ronco17} (see their Sec. 2.2.3), we compute the core mass of the collision remnant as the sum of the core masses of the target and the impactor. The final gaseous envelope is computed following \citet{Inamdar15}, who calculated the global atmospheric mass-loss fraction for planets with masses in the range of the Super-Earths and Mini-Neptunes. Although this study does not provide mass-loss fractions for collisions with gas giant planets with extended atmospheres, we use the same results due to lack of works on the subject.

For simplicity and following the results of \citet{Ogi20}, who report only one or two giant impacts when accounting for N-body interactions, we allow only one collision per planet, but compute all the possible results of that collision considering that all the less-massive planets in the same disc (with final periods $<$ 100 days), can be the impactor.
We compute mean values for the core mass, envelope mass and core ice fraction for each "family of impacts".The percentage of the envelope mass-loss due to impacts ranges between 11\% to 100\% with a mean of 55\%. If, for each family of impacts we consider the most destructive one (the one that generates the maximum envelope mass-loss), the percentage of the envelope mass-loss ranges between 16\% to 100\%, with a mean value of ~72\% for this latter case. Overall, collisions could reduce the mass of the envelope by a factor of $\sim$2.

After computing collisions, we compute the mass-loss due to photoevaporation (only with Model B) for the mean and maximum values of each family of impacts. In Fig.\ref{fig1_AppD}, which is similar to our previous Fig. \ref{fig_MR}, we compare the planet population affected only by photoevaporation (as in Fig. \ref{fig_MR}, green diamonds) with the planet population that also suffered a collision after gas dissipation (colored-circles). The color-bar of the circles represents the final water mass fraction with respect to the total heavy-element content \footnote{It is important to remark that due the simplification of the chemistry in our disc model, `water' and `ice' refer throughout this work to all species with condensation temperatures below 170 K.}, after collisions and photoevaporation are calculated. The black and yellow circles' borders represent those planets with mean and maximum envelope mass-loss after collisions, respectively; followed by envelope mass-loss due to photoevaporation. The grey border circles denote the naked cores of the planets that lost their entire atmosphere either after the collision, or after the collision followed by photoevaporation. For the cores resulting nude after the collision, the radii is computed following \citet{Zeng19} as in the main text.

The water mass fraction evidences some mixing of material due to collisions, fact that can be appreciated in all the figures of this Appendix. Indeed, approximately 33\% (20\%) of the resulting planets of the mean (maximum) collisional model have a final water mass fraction of $0.05\leq f_{\rm ice,f} <0.45$, compared to the $\sim$4\% when collisions were not considered (Sect.\ref{sec_results}). Nevertheless, most of the planets with this intermediate $f_{\rm ice,f}$, are still water-rich since they typically have $f_{\rm ice,f}>0.3$. This occurs because the original half-rock/half-water cores were more massive than the pure dry ones, and hence contribute with a non-negligible amount of water when colliding to pure rocky planets.
Moreover, 30\% of the resulting planets preserves its pure dry composition (for both collisional models); and 37\% (50\%) an $f_{\rm ice,f}\approx0.5$ for the mean (maximum) collisional models. This happens because many collision occur among cores that have originally the same composition.

There are some remarkable aspects to highlight from Fig.\ref{fig1_AppD}. First, the synthetic planets fill the delimiting M-R trends of the simulated planets of Fig. \ref{fig_MR}, accounting much better for the mean density diversity of real exoplanets. 
Second, bare rocky planets now can be as massive as 8 $\Mearth$ (compared to 5 $\Mearth$  without collisions), which fills better the exoplanets clustering around the Earth-like composition trend, which seems to extend until $\Mp\sim 10 \,\Mearth$. Finally, very energetic impacts are able to produce bare icy cores 
(grey-border blue circles of Fig.\ref{fig1_AppD}), which would explain the existence of a few exoplanets with $\Mp\sim 70 \,\Mearth$ and $\Rp \sim 4 \, \RE$. According to our formation-evolution model, such planets should be half-rock/half-water by mass. In addition, the ability of collisions to produce bare cores would move objects from $\Rp\sim8 \,\RE$ to lower radii. This is better appreciated in Fig.\ref{fig2_AppD}, where we repeat the histograms of the size distributions from Fig.\ref{fig_histEvap} for the two collisional models. From this histograms it is clear that the model considering the maximum amount of H-He removed by collisions gives the best match with observations: the valley in planet radii occurs at $\Rp\approx 1.8-2.1 \,\RE \,$ and the peaks at 1.4 and 2.8 $\RE$, giving a better agreement with the latest estimates \citep{Martinez19, VanEylen18} than the pure bare cores of Fig.\ref{fig_hist_cores}, whose peaks matched \citet{Fulton17} fairy well. 
Overall, the inclusion of a few giant impacts seems to be a crucial process to better reproduce the size and mass of short period exoplanets.

\end{appendix}


\end{document}